\documentclass[acus]{JAC2000}

\usepackage{graphicx}
\usepackage{citesort}       
\usepackage[]{amsmath}
\usepackage[]{amssymb}

\begin{document}
\title{COMPRESSION OF HIGH-CHARGE ELECTRON BUNCHES\thanks{Work supported in part by Fermilab 
which is operated 
by URA, Inc.\ for the U.S. DoE under contract
\mbox{DE-AC02-76CH03000}. 
\mbox{e-mail}:~\mbox{mjfitch@pas.rochester.edu}}}

\author{M. J. Fitch, A. C. Melissinos; University of Rochester, Rochester NY 14627, USA\\
N. Barov, J.-P. Carneiro, H. T. Edwards, W. H. Hartung; FNAL, 
Batavia IL 60510, USA}

\maketitle

\begin{abstract} 
The A{\O} Photoinjector at Fermilab can produce high charge (10-14 nC) 
electron bunches of low emittance ($20\pi$~mm-mrad for 12 nC). We have 
undertaken a study of the optimal compression conditions. Off-crest 
acceleration in the 9-cell capture cavity induces an energy-time 
correlation, which is rotated by the compressor chicane (4 dipoles).
The bunch length is measured using streak camera images of optical 
transition radiation. We present measurements 
under various conditions, including the effect of the laser pulse length
(2 ps sigma Gaussian vs.\ 10 ps FWHM flat top). The best compression to 
date is for a 13.2 nC bunch with $\sigma_{z}=0.63$~mm (1.89~ps), which 
corresponds to a peak current of 2.8~kA.
\end{abstract}

\section{INTRODUCTION}

Electron beams with short bunch lengths are desirable for 
high energy physics, free-electron lasers, and other 
applications.

In this paper we report on studies of compression
at the A{\O} Photoinjector of Fermilab 
\cite{Colby:1997,Carneiro:1999} with a chicane 
of four dipoles as measured by a picosecond streak camera.
The photoinjector was prototyped for the TeSLA Test Facility \cite{Edwards:1995}, and
in that context, there are three 
stages of acceleration and compression, and the chicane is the first 
compressor. The gun is a 1.625-cell $\pi$-mode normal conducting 
copper structure at 1.3~GHz whose backplane accepts a molybdenum plug 
coated with a Cs$_{2}$Te photocathode. Solenoids for emittance 
compensation surround the gun. A superconducting Nb nine-cell cavity 
accelerates the beam to 16--18~MeV. After the dipole chicane are 
experimental and diagnostic beamlines.

These streak 
camera measurements support electro-optic sampling measurements 
reported in a companion paper in these proceedings. Emittance 
measurements are reported by J.-P. Carneiro \emph{et al}.\ in these 
proceedings, and for 12~nC the normalized emittance is 
$\epsilon_{n}=20\pi$~mm-mrad without compression. The issue of 
emittance growth during bunch compression
\cite{Carlsten:1995,Carlsten:1996,CarlstenRussell:1996,Dohlus:1996,Braun:2000}
is under study, though preliminary studies suggest an emittance 
increase of approximately a factor 
of two.

\section{EXPERIMENT}

The streak camera is a Hamamatsu C5680-21S 
streak camera with M5676 fast sweep module\footnote{We thank A.~Hahn, 
FNAL Beams Division, for 
the loan of the streak camera.} read out by a Pulnix progessive scan 
digital CCD camera. Calibration was done using a short UV 
laser pulse and a thick fused silica delay block. We find a 
calibration of 3.9~pixels/ps at the fastest sweep speed with a 
limiting resolution of about 1~ps. After one year, the calibration was 
repeated, and the high-voltage sweep of the streak tube had degraded 
somewhat to 3.6~pixels/ps, so we report at most a 10\%  systematic 
uncertainty in the bunch length measurements.

Optical transition radiation (OTR) light \cite{Happek:1991,Shibata:1994,Lai:1994}
was imaged by all-reflective 
optics to the slit of the streak camera. 
OTR is prompt,
and has a characteristic opening angle of $1/\gamma$, and in 
our case $E=\gamma mc^{2}\sim 16$~MeV, so 
$1/\gamma\sim$~32~mrad or 1.8$^{\circ}$.
An out-of-plane (periscope) 
bend rotates the image so that the vertical direction of the beam 
falls on the horizontal slit. This is desirable to diagnose 
aberrations since the chicane bends in the vertical plane.

The photocathode drive laser (built by the University of Rochester) 
is a lamp-pumped Nd:glass system frequency-quadrupled to the UV 
($\lambda=263$~nm) \cite{Fry:1999}.  The UV 
laser pulse is a Gaussian with $\sigma_{t}=1.9$~ps.  Then, the UV pulses 
are temporally shaped to an
approximate flat-top distribution with 10.7~ps FWHM. We have measured 
the bunch length with both the long and the short laser pulse.

A number of streak images was acquired at each setting. After 
subtracting a constant background trace from the projected image,
(unsubtracted noise which varied from image to 
image was within $\pm 0.5$~pixel.) each streak image trace
is fit to a Gaussian. 
The mean value from the ensemble at each setting is reported, with
 error bars assigned from the statistical spread of values from this 
ensemble. The exception to this is in Figure \ref{fig:MOB16-1}, where 
each streak image is correlated with the charge measured on that shot, 
and the error bar is the error in the 
Gaussian fit.

\section{RESULTS: LONG LASER PULSE}

\begin{figure}[htb]
\centering
\includegraphics*[width=65mm]{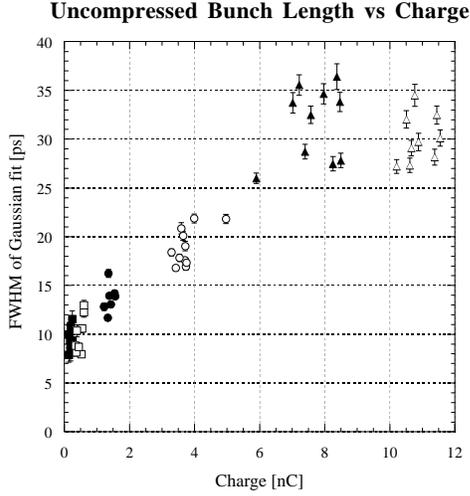}
\caption{Uncompressed bunch length vs.\ charge for the long (10~ps 
    FWHM) laser pulse length. }
\label{fig:MOB16-1}
\end{figure}

\begin{figure}[htb]
\centering
\includegraphics*[width=65mm]{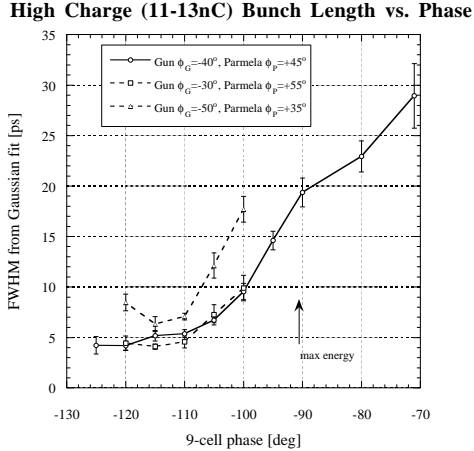}
\caption{Compression 
    vs.\ Phases of the 9-cell cavity and gun for the 10~ps laser pulse 
    length. The width of the focus mode image was 2.77~ps FWHM, and 
    the data were corrected assuming this broadening adds in 
    quadrature to the real width.}
\label{fig:MOB16-2}
\end{figure}

In Figure \ref{fig:MOB16-1} we give the uncompressed bunch length 
versus charge. At low charge, the bunch length is the same as that of 
the UV laser pulse on the cathode, but increases dramatically at 
higher charges. At high charge, ($\sim$11--13~nC), the beam was compressed
and measured as a function of the accelerating phases (Figure 
\ref{fig:MOB16-2}).
We set the middle pair of dipole chicane magnets to the nominal values (current 
$I=+2.0$~A or 680~Gauss), and reduce the outer pair slightly for 
vertical steering (typically $-1.9$ to $-1.95$~A). The phases of the gun 
RF and 9-cell cavity RF are recorded as the ``set phase'' from the 
control system ({\sc unix}). In addition to the set phase, we give 
the 9-cell phase for maximum energy (crest). The gun phase is referenced 
to {\sc parmela} by the curve of charge transmission 
vs.\ gun phase.

The point of best compression (Figure 
\ref{fig:MOB16-2}) is not sensitive to the gun phase, 
however the bunch lengthens if the phase is too early. Even for high 
charge, the measured bunch length is easily compressed to less than 1~mm $\sigma_{z}$
(or 3~ps), and the optimum is 0.63~mm $\sigma_{z}$ 
(1.89~ps).

\begin{figure}[htb]
\centering
\includegraphics*[width=65mm]{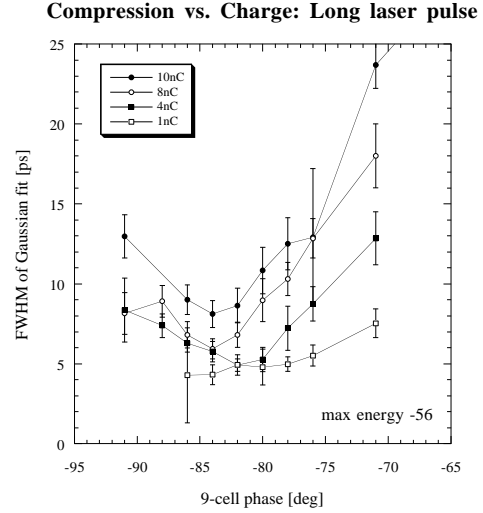}
\caption{Compression 
    vs.\ charge for the 10~ps laser pulse length.}
\label{fig:MOB16-3}
\end{figure}

Repeating this experiment as a function of charge,  we find that the 
minimum bunch length is shorter at lower charge (Figure 
\ref{fig:MOB16-3}), because
non-linear space charge growth is uncompensated.
The phase of optimal compression is only weakly dependent on the 
charge, shifting by 4$^{\circ}$ from 1~nC to 10~nC.

\section{RESULTS: SHORT LASER PULSE}

The compression experiments were repeated with a short Gaussian laser 
pulse ($\sigma_{t}=2$~ps) on the cathode, 
with the expectation that the space charge growth 
of the bunch length would be more 
severe. 

\begin{figure}[htb]
\centering
\includegraphics*[width=65mm]{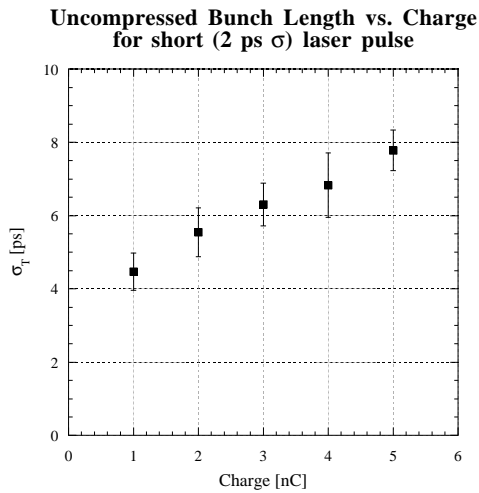}
\caption{Uncompressed bunch length 
    vs.\ charge for the 2~ps laser pulse length.}
\label{fig:MOB16-4}
\end{figure}

\begin{figure}[htb]
\centering
\includegraphics*[width=65mm]{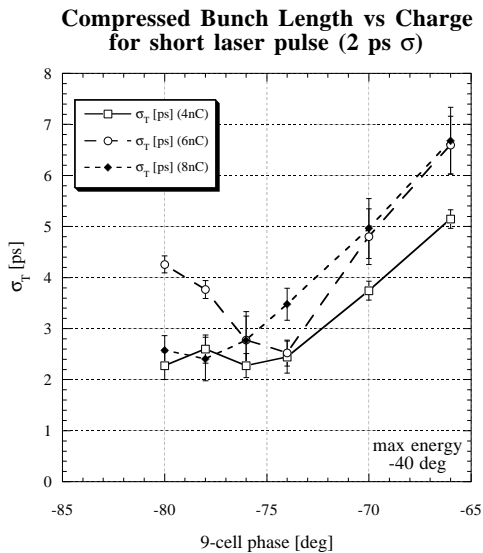}
\caption{Compressed bunch length 
    vs.\ charge for the 2~ps laser pulse length.}
\label{fig:MOB16-5}
\end{figure}

With the chicane dipole magnets off and degaussed, we measured the 
uncompressed bunch length with a streak camera looking at OTR 
radiation as before (Figure \ref{fig:MOB16-4}) from 1~nC to 5~nC. 
Even at 
low charge, the bunch length is more than a factor of two
longer than the initial laser pulse length on the cathode, and 
increases linearly with charge. The compressed bunch 
length for the short laser pulse is shown in Figure \ref{fig:MOB16-5},
and the minimum is slightly larger, and at a larger angle off-crest.

\section{PEAK CURRENT}

One figure of merit for (sub)picosecond electron bunches is the peak 
current, which depends on both the charge and the bunch length. If the 
beam is Gaussian in time,
\begin{equation}
    I(t)=\frac{Q}{\sqrt{2\pi}\,\sigma_{t}} \, \exp(-t^{2}/(2 \sigma_{t}^{2}))
    \label{eq:gaus-i-t}
\end{equation}
Then the peak current is by definition the peak value of the current profile:
\begin{equation}
    I_{p} = \frac{Q}{\sqrt{2\pi}\,\sigma_{t}} =\frac{2\sqrt{2\ln 
    2}\,Q}{\sqrt{2\pi}\,\tau}
    \label{eq:Ipeak-sig} 
\end{equation}
for the rms bunch length $\sigma_{z}$ or the full width at half maximum 
(FWHM) $\tau$.
 
We know of two other facilities which report peak current at or above 2~kA 
which are 1.97~kA at the AWA \cite{Conde:1998} and 2.3~kA 
at the CLIC Test 
Facility (CTF-II) at CERN \cite{Braun:2000}. 
Our reported best peak current of 2.8~kA is a significant improvement.

\bibliographystyle{unsrt}
\bibliography{MOB16}

\end{document}